\documentclass[aps,prl,reprint,floatfix,nobalancelastpage,nofootinbib,superscriptaddress]{revtex4-2}

\usepackage{amsmath,amssymb,mathtools}
\usepackage{bm}
\usepackage{braket}
\usepackage{slashed}
\usepackage{dsfont}
\usepackage{booktabs}
\usepackage{array}
\usepackage{graphicx}

\usepackage{xcolor}
\definecolor{nicered}{rgb}{0.5,0.,0.}
\definecolor{nicegreen}{rgb}{0.,0.5,0.}
\definecolor{niceblue}{rgb}{0.,0.,0.5}
\usepackage[colorlinks,citecolor=nicegreen,linkcolor=nicered,urlcolor=niceblue]{hyperref}
\usepackage{orcidlink}

\newcommand{\dd}{\mathrm{d}}
\newcommand{\ii}{\mathrm{i}}
\newcommand{\ee}{e}

\newcommand{\Mcal}{\mathcal{M}}

\newcommand{\Ocal}{\mathcal{O}}
\newcommand{\Ical}{\mathcal{I}}
\newcommand{\Tr}{\operatorname{Tr}}

\newcommand{\Real}{\operatorname{Re}}
\newcommand{\Imag}{\operatorname{Im}}
\newcommand{\GeV}{\mathrm{GeV}}
\newcommand{\dt}{\widetilde d_\tau^W}
\newcommand{\dg}{\widetilde d_\tau^\gamma}

\newcommand{\ecm}{e\,\mathrm{cm}}

\begin{document}

\title{Transverse Tau Spin Correlations and the Weak Electric Dipole Moment
\texorpdfstring{\\}{ } at Future \texorpdfstring{\(Z\)}{Z} Factories}

\author {Xin-Yu Du\orcidlink{0009-0005-6925-9824}}\email{2020dxy@sjtu.edu.cn}
\author {Zi-Yue Zou\orcidlink{0009-0004-4446-527X}}\email{ziy\_zou@sjtu.edu.cn}
\author {Xiao-Gang He\orcidlink{0000-0001-7059-6311}}\email{hexg@sjtu.edu.cn}
\author {Keping Xie\orcidlink{0000-0003-4261-3393}}\email[Corresponding author: ]{kepingxie@sjtu.edu.cn}
\affiliation{State Key Laboratory of Dark Matter Physics, Key Laboratory for Particle Astrophysics and Cosmology (MOE) \& Shanghai Key Laboratory for Particle Physics and Cosmology, Tsung-Dao Lee Institute \& School of Physics and Astronomy, Shanghai Jiao Tong University, Shanghai 200240, China\looseness=-1}

\date{September 16, 2026}

\begin{abstract}
Tau spin correlations at the \(Z\) pole probe CP violation with a
sensitivity that depends strongly on the spin projection. For a fixed \(Z\) sample, transverse
moments avoid the chirality suppression of longitudinal and triple-product
observables, enhancing the response by a factor \(m_Z/2m_\tau=26\). Two transverse moments also select the
axial tau charge, gaining \(|a_\tau/v_\tau|=13.5\); the sign-weighted
normal moment has a response \(521\) times that of the triple product.
These two transverse moments give single-parameter uncertainties no more than \(14\%\) above our calculated Cram\'er--Rao bound for independent single-pion measurements, approaching the best precision within this analysis.
A five-moment fit with \(5\%\) off-peak luminosity yields statistical
uncertainties of \(4.5\,(5.5)\times10^{-7}\) and
\(2.7\,(3.2)\times10^{-6}\) on \(\Real\dt\) and \(\Imag\dt\), respectively,
at FCC-ee (CEPC), improving on ALEPH by three orders of magnitude.
The electromagnetic-dipole uncertainty is \(9.6\,(12)\times10^{-6}\) on \(\Real\dg\), two orders beyond Belle. These projections assume ideal angular reconstruction and constant form factors across the scan.
\end{abstract}

\maketitle

\textit{Introduction.}---In the Standard Model (SM), established
CP violation originates in the Cabibbo--Kobayashi--Maskawa (CKM) phase.
Its contribution to electroweak baryogenesis is insufficient to explain the
observed baryon asymmetry~\cite{Huet:1994jb}. Permanent electric
dipole moments provide sensitive null tests for additional CP-violating
phases~\cite{Roussy:2022cmp,Chupp:2017rkp}.
The tau dipole is the least constrained among charged leptons: the tau's
short lifetime precludes storage, so its dipole must be inferred from
correlated spins in pair production and decay~\cite{Buttazzo:2026amk}.
At the \(Z\) pole, FCC-ee and CEPC will each deliver order
\(10^{12}\) \(Z\) bosons~\cite{FCC:2025lpp,CEPCStudyGroup:2023quu,Ai:2024nmn},
a million times ALEPH's sample~\cite{ALEPH:2002kbp}.

The tau weak electric dipole moment \(d_\tau^W\), the CP-odd
\(Z\tau\tau\) form factor, first arises at four loops in the SM and is
negligible at foreseeable experimental sensitivities~\cite{Kosnik:2026wdm,Pospelov:2013sca}.
An observable signal would therefore establish new physics.
The weak and electromagnetic dipoles enter the tau vertices as
\begin{align}
  \Gamma_\tau^{Z,\nu}&=\ii\ee\Big[v_\tau\gamma^\nu-a_\tau\gamma^\nu\gamma^5
    +\tfrac{\dt}{2m_\tau}\gamma^5\sigma^{\nu\rho}q_\rho\Big],
  \label{eq:vertexZ}\\
  \Gamma_\tau^{\gamma,\nu}&=\ii\ee\Big[Q_\tau\gamma^\nu
    +\tfrac{\dg}{2m_\tau}\gamma^5\sigma^{\nu\rho}q_\rho\Big],
  \label{eq:vertexG}
\end{align}
Here \(q\) is the boson momentum. We use the dimensionless normalization
\(d_\tau^W=\ee\dt/(2m_\tau)\).

The established method~\cite{Bernreuther:1989kc,Bernreuther:1993nd}
uses parity-violating tau decays as spin analyzers: CP-odd correlations
of decay-product directions respond linearly to the dipole.
ALEPH obtained~\cite{ALEPH:2002kbp}
\begin{equation}
  |\Real\dt|<0.91\times10^{-3},\qquad |\Imag\dt|<2.01\times10^{-3}
  \label{eq:aleph}
\end{equation}
at \(95\%\) confidence level (CL). The absorptive component is
independently observable~\cite{Huang:2025ghw,Du:2026yoi}.

A spin selection rule controls the sensitivity. The SM current
aligns both tau spins with the production axis; the longitudinal and
triple-product moments require a helicity flip and carry a factor
\(2m_\tau/m_Z\simeq1/26\). Transverse polarization observables~\cite{Vidal:1998jc} avoid this flip and retain the enhancement
\(m_Z/2m_\tau\); two moments also select the axial charge \(a_\tau\), larger in
magnitude than \(v_\tau\) by \(13.5\). We explain this hierarchy and its
growth with \(\sqrt s/2m_\tau\). A joint fit with
\(5\%\) of the \(Z\)-run luminosity off peak separates the weak and electromagnetic
dipoles and improves the timelike photon-dipole precision tenfold.

\textit{Framework.}---We describe
\(e^-(p_1)e^+(p_2)\to\gamma^*,Z\to\tau^-(k_1)\tau^+(k_2)\) using the vertices
in Eqs.~\eqref{eq:vertexZ}--\eqref{eq:vertexG}
and the electron couplings \(v_e,a_e\) and \(Q_e=-1\).
We set the anomalous magnetic dipoles to zero: their CP-even contributions
to CP-odd means vanish at linear order, as verified numerically to relative
precision \(10^{-14}\).
An electron CP-odd dipole flips chirality and produces same-helicity
initial states absent from the SM current. Its interference with the SM
therefore vanishes; corrections enter at \(\Ocal(m_e,d_e^2)\).

The small leptonic vector charge \(v_\ell\)\(\propto-\tfrac12+2s_W^2\)
is sensitive to the electroweak input scheme.
The inputs \((\alpha,m_W,m_Z)\) fix both charges at tree level and give
\(A_\tau\equiv2v_\tau a_\tau/(v_\tau^2+a_\tau^2)=0.212\), the asymmetry
parameter associated with the angle-averaged tau polarization at the pole.
This differs from the measured
\(A_\tau^{\rm exp}=0.1439\pm0.0043\)~\cite{ALEPH:2005ab}. We therefore
retain the on-shell normalization for \(a_\ell\) and determine \(v_\ell/a_\ell\)
from the measured effective leptonic angle
\(s_{\rm eff}^2=0.23155\)~\cite{ALEPH:2005ab,ParticleDataGroup:2024cfk}:
\begin{equation}
  v_\ell=-0.04431,\qquad a_\ell=-0.60039 ,
  \label{eq:couplings}
\end{equation}
equal for \(\ell=e,\tau\) by universality. These give \(A_\tau=0.1468\).
The effective-angle measurement determines the ratio \(a_\tau/v_\tau\)
used below to a relative precision of \(1\%\).

The unnormalized production density matrix is
\(R_{FF'}=\tfrac14\sum_\lambda\Mcal_{F\lambda}\Mcal^*_{F'\lambda}\), where
\(F=(h_-,h_+)\) labels the tau helicities. The two-body decays \(\tau^\mp\to\pi^\mp\nu\) act as
local spin analyzers~\cite{Tsai:1971vv},
\(D_\mp=\tfrac12(I_2\pm\alpha_\mp\hat\ell^{\,\mp}\!\cdot\bm\sigma)\)
with \(\alpha_\mp=1\). The visible weight is
\(W=\Tr[R\,(D_-\otimes D_+)]\).

\textit{Angular moments and responses.}---We quantize each tau spin along its
helicity axis and express the pion directions in their respective parent
rest frames using a common right-handed triad,
\begin{equation}
  \hat z=\hat k_-,\qquad
  \hat y=\frac{\hat p_1\times\hat k_-}{|\hat p_1\times\hat k_-|},\qquad
  \hat x=\hat y\times\hat z,
  \label{eq:triad}
\end{equation}
so that
\(\hat\ell^{\,\mp}=(\sin\theta_\mp\cos\phi_\mp,\sin\theta_\mp\sin\phi_\mp,
\cos\theta_\mp)\) for \emph{both} charges.  Under CP,
\(\hat z,\hat y,\hat x\) are invariant while
\(\hat\ell^{\,\mp}\to-\hat\ell^{\,\pm}\), which fixes the CP assignments.
Because Eq.~\eqref{eq:triad} is not the \(\tau^+\) helicity frame, \(D_+\) is
evaluated with \(\theta_+\to\pi-\theta_+\), \(\phi_+\to-\phi_+\); omitting this conversion spuriously removes the triple product.  Here, \(\theta\)
without a subscript denotes the production angle,
\(\cos\theta=\hat k_-\!\cdot\hat p_1\).

The single-spin longitudinal asymmetry and triple product of
Bernreuther and Nachtmann~\cite{Bernreuther:1989kc,Bernreuther:1993nd},
studied further in Refs.~\cite{Bernabeu:2004ww,Bernabeu:2006wf,Sun:2024vcd},
project onto \(\hat z\).
Here, transverse includes the in-plane and normal directions perpendicular to the tau flight axis.
Three transverse moments complete the set, including the sign-weighted normal moment~\(\Ocal_4\):

\begin{align}
  \Ocal_1&=(\hat\ell^{\,-}+\hat\ell^{\,+})\cdot\hat z
          =\cos\theta_-+\cos\theta_+,
  \label{eq:O1}\\
  \Ocal_2&=(\hat\ell^{\,-}\times\hat\ell^{\,+})\cdot\hat z
          =\sin\theta_-\sin\theta_+\sin(\phi_+-\phi_-),
  \label{eq:O2}\\
  \Ocal_3&=(\hat\ell^{\,-}+\hat\ell^{\,+})\cdot\hat x
          =\sin\theta_-\cos\phi_-+\sin\theta_+\cos\phi_+,
  \label{eq:O3}\\
  \Ocal_4&=\mathrm{sgn}(\cos\theta)\,
   (\hat\ell^{\,-}+\hat\ell^{\,+})\cdot\hat y
          =\mathrm{sgn}(\cos\theta)\,\Ocal_5 ,
  \label{eq:O4}\\
  \Ocal_5&=(\hat\ell^{\,-}+\hat\ell^{\,+})\cdot\hat y
          =\sin\theta_-\sin\phi_-+\sin\theta_+\sin\phi_+ .
  \label{eq:O5}
\end{align}

Here \(\beta=\sqrt{1-4m_\tau^2/m_Z^2}=0.99924\) is the tau velocity, and
\(\Delta_0\equiv2a_\tau^2\beta^2+v_\tau^2(3-\beta^2)=0.72378\)
is the spin- and angle-summed tree-level normalization. Since
\(\Gamma(Z\to\tau\tau)\propto\beta\Delta_0\), every response carries
\(\Delta_0^{-1}\). The electron asymmetry parameter is
\(A_e=2v_ea_e/(v_e^2+a_e^2)=0.1468\). To linear order in the dipoles,
\begin{align}
  \langle\Ocal_1\rangle&=-\frac43\frac{\beta v_\tau}{\Delta_0}\,
    \Imag\big[\dt+\ii\epsilon_v\dg\big] ,
  \label{eq:O1resp}\\
  \langle\Ocal_2\rangle&=-\frac49\frac{\beta v_\tau}{\Delta_0}\,
    \Real\big[\dt+\ii\epsilon_v\dg\big] ,
  \label{eq:O2resp}\\
  \langle\Ocal_3\rangle&=-\frac{\pi}{2}\frac{m_Z}{2m_\tau}
    \frac{a_\tau A_e\beta^2}{\Delta_0}\,
    \Imag\big[\dt+\ii\epsilon_a\dg\big] ,
  \label{eq:O3resp}\\
  \langle\Ocal_4\rangle&=-\frac23\frac{m_Z}{2m_\tau}
    \frac{a_\tau\beta^2}{\Delta_0}\,
    \Real\big[\dt+\ii\epsilon_v\dg\big] ,
  \label{eq:O4resp}\\
  \langle\Ocal_5\rangle&=-\frac{\pi}{2}\frac{m_Z}{2m_\tau}
    \frac{v_\tau A_e\beta}{\Delta_0}
    \big(\Real\big[\dt+\ii\epsilon_a\dg\big]+\xi\,\Imag\dt\big) .
  \label{eq:O5resp}
\end{align}

At the pole, the propagator ratio is purely imaginary,
Eq.~\eqref{eq:propratio}, so \(\dg\) enters each observable as an
imaginary shift of \(\dt\), with
\begin{equation}
\begin{aligned}
  \epsilon_v&=\frac{\Gamma_Z}{m_Z}\frac{Q_ev_e}{v_e^2+a_e^2}=3.35\times10^{-3},\\
  \epsilon_a&=\frac{\Gamma_Z}{m_Z}\frac{Q_e}{2v_e}=0.309 .
\end{aligned}
  \label{eq:eps}
\end{equation}
Only the weak term in \(\Ocal_3\)
contains \(A_e\). The ratio,
\(\epsilon_a/\epsilon_v=(v_e^2+a_e^2)/2v_e^2=92.3\), makes the relative
photon contribution to \(\Ocal_4\) \(92\) times smaller than to
\(\Ocal_3\). \(\Ocal_5\) has no pure-\(Z\) sensitivity to
\(\Imag\dt\): the final term in Eq.~\eqref{eq:O5resp} arises from
SM-photon interference with the \(Z\) dipole, with coefficient
\begin{equation}
  \xi=\frac{\Gamma_Z}{m_Z}\frac{Q_eQ_\tau}{v_\tau A_ea_e}=7.0 .
  \label{eq:xi}
\end{equation}
This reflects propagator suppression \(\Gamma_Z/m_Z=0.027\)
without the small charge product \(v_\tau A_ea_e=3.9\times10^{-3}\).
Equations~\eqref{eq:O1resp}--\eqref{eq:O5resp} retain terms through first order in
\(D^\gamma/D^Z\) and omit the electron's photon current;
Eq.~\eqref{eq:contam} gives the full numerical responses.

For the pure-Z weak-dipole terms, the \(\dt\) component follows from naive time
reversal, \(\hat\ell^{\,\mp}\to-\hat\ell^{\,\mp}\) and
\(\hat k_-\to-\hat k_-\): \(\Ocal_1,\Ocal_3\) are even and probe
\(\Imag\dt\), whereas \(\Ocal_2,\Ocal_4,\Ocal_5\) are odd and probe
\(\Real\dt\). Among the first four moments, only \(\Ocal_3\) involves electron charges,
through \(A_e\); \(\Ocal_1\), \(\Ocal_2\), and \(\Ocal_4\) probe
the \(\tau\) sector alone. \(\Ocal_4\) contains no vector charge
and is therefore largely insensitive to the input scheme: replacing
\(s^2_{\rm eff}\) by the tree-level \(s_W^2=1-m_W^2/m_Z^2\)
multiplies responses by \(1.44\) for each vector charge.
This changes Eqs.~\eqref{eq:O1resp}--\eqref{eq:O3resp} by \(44\%\)
and Eq.~\eqref{eq:O5resp}, with two vector charges, by \(108\%\),
but Eq.~\eqref{eq:O4resp} by only \(0.6\%\).

Two factors enhance the responses. The dipole vertex factor
\(q_\rho/2m_\tau\), with \(q^2=m_Z^2\), suggests an enhancement
\(m_Z/2m_\tau=25.66\), whose square enters the rate response
\(\sigma/\sigma_{\rm SM}=1+A|\dt|^2\), \(A=m_Z^2\beta^2/(4m_\tau^2\Delta_0)
=908.2\). It survives in Eqs.~\eqref{eq:O3resp} and
\eqref{eq:O4resp}, but cancels in Eqs.~\eqref{eq:O1resp} and
\eqref{eq:O2resp}: the terms for \(\Ocal_1\) and \(\Ocal_2\)
contain the compensating factor \(\sqrt{1-\beta^2}=2m_\tau/m_Z\).

In the \(e^+e^-\) rest frame, \(q^\mu=(m_Z,\bm 0)\), the dipole
operator becomes
\(\gamma^5\sigma^{i\rho}q_\rho/2m_\tau=-\ii(m_Z/2m_\tau)\,\Sigma^i\),
an enhanced pure spin operator. Here \(\Sigma^i=\mathrm{diag}(\sigma^i,\sigma^i)\) is the Dirac spin
matrix. An electric dipole couples as
\(\bm d\cdot\bm S\), rotating polarization at fixed magnitude.
At \(\beta\to1\), the chirality-conserving SM current predominantly
produces \((h_-,h_+)=(+,-)\) or \((-,+)\), with spins along
\(\pm\hat z\); \(S_z=0\) requires a chirality flip, \(m_\tau/E\).
The dipole thus \emph{tilts the spins away from \(\hat z\)}.
Since \(\Ocal_1\) and \(\Ocal_2\) use only \(\hat z\), their
interference requires the helicity-flip amplitude. Transverse moments
\(\Ocal_3\), \(\Ocal_4\) and \(\Ocal_5\) measure the tilt
linearly and retain the \(m_Z/2m_\tau\) enhancement.

Couplings provide the second factor:
Eqs.~\eqref{eq:O1resp} and \eqref{eq:O2resp} carry the tau
\emph{vector} charge \(v_\tau\), whereas Eqs.~\eqref{eq:O3resp} and
\eqref{eq:O4resp} carry the \emph{axial} charge \(a_\tau\).
Small \(v_\ell\) suppresses \(\Ocal_1\) and \(\Ocal_2\) by
\(|a_\tau/v_\tau|=1/(1-4s^2_{\rm eff})=13.5\).

Consequently, \(\Ocal_3\) probes the same
\(\Imag\dt\) as \(\Ocal_1\) using the same events, with a response
larger by
\begin{equation}
  \frac{3\pi}{8}\frac{m_Z}{2m_\tau}\frac{\beta a_\tau A_e}{v_\tau}=60.1 ,
  \label{eq:O3-over-O1}
\end{equation}
where \(A_e=0.1468\) dilutes the axial factor because \(\Ocal_3\)
takes its transverse axis from the beam.

Unlike the moments in Eqs.~\eqref{eq:O1}--\eqref{eq:O2},
\(\Ocal_4\) uses the production-angle weight to retain the full
axial contribution. Its response exceeds that of \(\Ocal_2\) by
\begin{equation}
  \frac32\frac{m_Z}{2m_\tau}\frac{\beta|a_\tau|}{|v_\tau|}=521
  =25.7\times13.5\times1.50,
  \label{eq:gain}
\end{equation}
where the three factors correspond to removing the helicity suppression,
replacing \(v_\tau\) with \(a_\tau\), and angular integration.

The weight \(\mathrm{sgn}(\cos\theta)\) in Eq.~\eqref{eq:O4} is
essential: the axial term is odd in \(\cos\theta\) and cancels under
the symmetric integration implicit in \(\Ocal_1\) and \(\Ocal_2\).
The smooth alternative \(\cos\theta\) is \(7\%\) less efficient.

Single-pion projections have equal second moments,
\(\langle(\hat\ell^{\,\mp}\!\cdot\hat x)^2\rangle
=\langle(\hat\ell^{\,\mp}\!\cdot\hat y)^2\rangle
=\langle(\hat\ell^{\,\mp}\!\cdot\hat z)^2\rangle=\tfrac13\).
Like \(\Ocal_1\) and unlike \(\Ocal_2\), these transverse moments use
inclusive single-pion samples, larger than double-pion samples by
\(1/(B_\pi\epsilon)=11.6\).

\(\Ocal_1\) requires only the pion energy fraction, since
\(\cos\theta_-\simeq2E_\pi/E_\tau-1\). Transverse moments require
the pion azimuth about the tau flight direction, reconstructed from
the impact parameter and beam constraint, as in
Ref.~\cite{ALEPH:2002kbp}; see also Ref.~\cite{Yu:2019oal}.
Response gains of \(60\) to \(521\) motivate reconstruction;
detector studies must quantify signal dilution.

End Matter discusses identifiability under kinematic ambiguity.

\begin{figure*}[t]
\centering
\includegraphics[width=\textwidth]{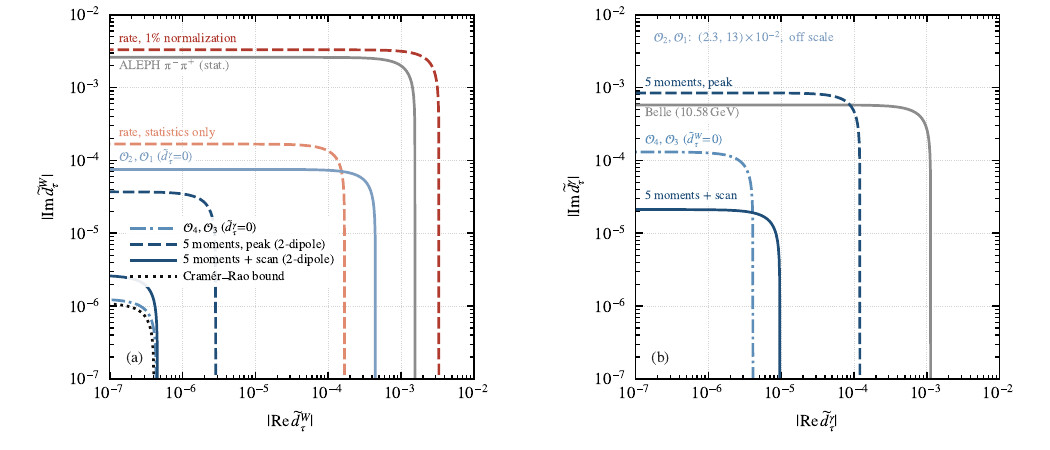}
\caption{\(1\sigma\) FCC-ee contours from
Table~\ref{tab:reach}, on logarithmic axes. Arms mark the tabulated marginal errors; correlations are set to zero for plotting.
(a) Complex \(\dt\) plane. The dark solid contour shows the joint-fit reach,
Eq.~\eqref{eq:final}; the rate contours compare the Poisson radius with the
normalization-limited radius \(\sqrt{\delta_N/A}\) for
\(\delta_N=1\%\). (b) Complex \(\dg\) plane at \(q^2=m_Z^2\),
Eq.~\eqref{eq:gammareach}. The \(5\)-moment contours show the two-dipole fit.
Off-peak data improve both photon-dipole uncertainties by more than an order of magnitude;
the \(\Ocal_2,\Ocal_1\) contour lies off scale, as indicated.}
\label{fig:reach}
\end{figure*}

\textit{Sensitivity reach.}---For \(N_Z=6.0\times10^{12}\) (\(4.1\times10^{12}\)) at
FCC-ee (CEPC), the effective sample sizes are
\(N_{\rm eff}^\mp=N_ZB_ZB_\pi\epsilon\) and
\(N_{\rm eff}^{+-}=N_ZB_ZB_\pi^2\epsilon^2\). With \(B_Z=0.033696\),
\(B_\pi=0.1082\)~\cite{ParticleDataGroup:2024cfk} and reference efficiency
\(\epsilon=0.8\) per pion side, the moment uncertainties are
\(\delta\langle\Ocal_{3,4}\rangle=[2/(3N_{\rm eff}^\mp)]^{1/2}\).
Inverting Eqs.~\eqref{eq:O3resp}--\eqref{eq:O4resp} gives
Table~\ref{tab:reach} and the \(1\sigma\) contours in
Fig.~\ref{fig:reach}.

\begin{table*}[t]
\centering
\caption{\(1\sigma\) uncertainties on the two dimensionless dipole
coefficients, compared with current bounds (Fig.~\ref{fig:reach}).
The first three FCC-ee and CEPC rows use full on-peak samples with the
\emph{other} dipole fixed to zero: the \(\dt\)
columns at \(\dg=0\), the \(\dg\) columns at \(\dt=0\), inferred
from the same moments through Eq.~\eqref{eq:eps}.
Cram\'er--Rao bounds are \((N\Ical_{ii})^{-1/2}\), using \(\Ical^{(1)}_{ii}\) from Eq.~\eqref{eq:fisher1} for
\(N=2N_{\rm eff}^\mp\) pion measurements treated independently at FCC-ee and CEPC,
and \(\Ical_{ii}\) from Eq.~\eqref{eq:fisherpipi} for the \(1901\)
ALEPH \(\pi^-\pi^+\) candidates; both computed for \(\dt\) only. The \(5\)-moment rows give the joint four-parameter fit to
Eqs.~\eqref{eq:O1}--\eqref{eq:O5}, with weak-dipole uncertainties marginalized over \(\dg\): ``peak'' uses
the full on-peak sample; ``\(+\) scan'' allocates \(5\%\) of the luminosity off peak,
giving the reach in Eqs.~\eqref{eq:final} and \eqref{eq:gammareach}. ALEPH errors are published double-pion statistical uncertainties. Belle probes the form factor at
\(\sqrt s=10.58\,\GeV\); the projections use squared momentum transfer \(m_Z^2\). Belle's \(95\%\) CL
intervals~\cite{Belle:2021ybo} give Gaussian \(\sigma\) as
half-width\(/1.96\), with central values consistent with zero
[Fig.~\ref{fig:reach}(b)].}
\label{tab:reach}
\begin{tabular}{llcccc}
\toprule
Sample & Estimator & \(\Real\dt\) & \(\Imag\dt\)
       & \(\Real\dg\) & \(\Imag\dg\)\\
\midrule
FCC-ee & \(\Ocal_2,\Ocal_1\)  & \(4.5\times10^{-4}\) & \(7.5\times10^{-5}\)
       & \(2.3\times10^{-2}\) & \(1.3\times10^{-1}\)\\
       & \(\Ocal_4,\Ocal_3\)  & \(4.4\times10^{-7}\) & \(1.3\times10^{-6}\)
       & \(4.1\times10^{-6}\) & \(1.3\times10^{-4}\)\\
       & Cram\'er--Rao        & \(4.1\times10^{-7}\) & \(1.1\times10^{-6}\)
       & --- & ---\\
       & 5 moments, peak & \(2.9\times10^{-6}\)
       & \(3.7\times10^{-5}\) & \(1.2\times10^{-4}\)
       & \(8.5\times10^{-4}\)\\
       & \(5\) moments \(+\) scan & \(4.5\times10^{-7}\) & \(2.7\times10^{-6}\)
       & \(9.6\times10^{-6}\) & \(2.1\times10^{-5}\)\\
\midrule
CEPC   & \(\Ocal_2,\Ocal_1\)  & \(5.4\times10^{-4}\) & \(9.2\times10^{-5}\)
       & \(2.7\times10^{-2}\) & \(1.6\times10^{-1}\)\\
       & \(\Ocal_4,\Ocal_3\)  & \(5.2\times10^{-7}\) & \(1.5\times10^{-6}\)
       & \(5.0\times10^{-6}\) & \(1.6\times10^{-4}\)\\
       & Cram\'er--Rao        & \(4.9\times10^{-7}\) & \(1.3\times10^{-6}\)
       & --- & ---\\
       & 5 moments, peak & \(3.5\times10^{-6}\)
       & \(4.5\times10^{-5}\) & \(1.5\times10^{-4}\)
       & \(1.0\times10^{-3}\)\\
       & \(5\) moments \(+\) scan & \(5.5\times10^{-7}\) & \(3.2\times10^{-6}\)
       & \(1.2\times10^{-5}\) & \(2.6\times10^{-5}\)\\
\midrule
ALEPH  & published stat.      & \(1.6\times10^{-3}\) & \(2.6\times10^{-3}\)
       & --- & ---\\
       & Cram\'er--Rao        & \(9.7\times10^{-4}\) & \(1.6\times10^{-3}\)
       & --- & ---\\
\midrule
Belle  & published limit & --- & ---
       & \(1.1\times10^{-3}\) & \(5.8\times10^{-4}\)\\
\bottomrule
\end{tabular}
\end{table*}

The hierarchy is stronger for the electromagnetic dipole. Each observable measures \(\dt+\ii\epsilon\dg\),
so its uncertainty on \(\dg\) is that on \(\dt\) divided by \(\epsilon\).
With the weak dipole fixed to zero, Eq.~\eqref{eq:eps} implies that \(\Ocal_3\)
alone yields an uncertainty of \(4.1\times10^{-6}\) on \(\Real\dg\),
a factor \(280\) below Belle's, while \(\Ocal_1\) and \(\Ocal_2\)
reach only \(10^{-2}\). The electromagnetic-dipole columns of
Table~\ref{tab:reach} are dominated by \(\Ocal_3\), with
\(\Ocal_4\) contributing to \(\Imag d_\tau^\gamma\): the reduced photon contamination of \(\Ocal_4\) also limits its photon-dipole sensitivity.

With the longitudinal and triple-product moments of
Bernreuther and Nachtmann~\cite{Bernreuther:1989kc,Bernreuther:1993nd},
FCC-ee gains factors of \(3.5\) to \(35\),
below the \(1.2\times10^3\) from \(\sqrt N\) scaling. The suppression \(2m_\tau/m_Z\) limits the gain from \(1.5\times10^6\) more events.
In single-parameter fits, transverse moments improve precision by \(3.6\times10^3\) on
\(\Real\dt\) and \(2.1\times10^3\) on \(\Imag\dt\); their uncertainties lie only \(8\%\) and
\(14\%\) above our calculated single-pion Cram\'er--Rao bound
\(\delta\theta_i\geq(N\Ical_{ii})^{-1/2}\) (End Matter).
This bound sets a lower uncertainty limit for unbiased estimators treating pion measurements independently and gauges how efficiently each moment extracts the single-pion angular information.
The transverse moments \(\Ocal_3\) and \(\Ocal_4\) thus approach the best statistical precision within this analysis.

A joint fit must separate the weak and electromagnetic contributions. At the pole, the propagators satisfy
\begin{equation}
  D^\gamma(m_Z^2)/D^Z(m_Z^2)=\ii\,\Gamma_Z/m_Z=0.0274\,\ii ,
  \label{eq:propratio}
\end{equation}
so photon exchange enters with a \(90^\circ\) relative phase.  On peak we
find
\begin{align}
  \langle\Ocal_3\rangle&=4.873\,\Imag\dt+1.505\,\Real\dg ,\\
  \langle\Ocal_4\rangle&=14.088\,\Real\dt-0.047\,\Imag\dg ,\\
  \langle\Ocal_5\rangle&=0.360\,\Real\dt+2.508\,\Imag\dt
                        -0.111\,\Imag\dg .
  \label{eq:contam}
\end{align}
The four-parameter fit to on-peak data is nearly degenerate: correlations
of \(-0.9994\) and \(+0.988\) increase the uncertainties to
\(2.9\times10^{-6}\) and \(3.7\times10^{-5}\).

\begin{figure}[t]
\centering
\includegraphics[width=\columnwidth]{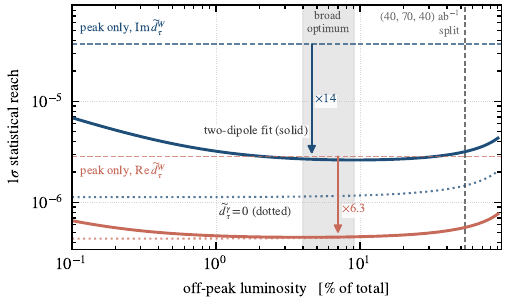}
\caption{Marginal \(1\sigma\) uncertainties on \(\dt\) from the five-moment
FCC-ee fit to weak and electromagnetic dipoles, versus the off-peak
luminosity fraction. This luminosity is shared equally among four energies,
\(m_Z\pm1.7,\pm3.5~\GeV\); at \(5\%\) off-peak luminosity, \(1.2\%\) of the tau
pairs are produced off peak. Shading marks the optimum; the vertical line denotes the
\((40,70,40)\,\mathrm{ab}^{-1}\) luminosity split~\cite{Janot:2015gjr}.
Solid: \(\dg\) fitted and marginalized over. Dotted:
\(\dg\) fixed to zero. Dashed: the peak-only fit with \(\dg\) free.
Arrows compare peak-only and scan uncertainties at the \(5\%\) benchmark.}
\label{fig:scan}
\end{figure}

Away from the peak, the degeneracy lifts because
\(D^\gamma/D^Z=(s-m_Z^2+\ii m_Z\Gamma_Z)/s\) acquires a real part that changes
sign across the resonance: at \(\sqrt s=87.7,\ 91.2,\ 94.7~\GeV\),
\begin{equation}
  \frac{D^\gamma}{D^Z}=-0.081+0.030\,\ii,\quad
  0.027\,\ii,\quad 0.073+0.025\,\ii ,
  \label{eq:offpeak}
\end{equation}
so the photon-induced terms reverse while the \(Z\)-dipole terms do not.
The fit uses all five moments. Both \(\Ocal_3\) and \(\Ocal_5\) probe
\(\Imag\dt\), Eq.~\eqref{eq:contam}, through different combinations: the first includes \(\Real\dg\), whereas the second does not. This breaks the absorptive degeneracy;
at \(5\%\) off peak, omitting \(\Ocal_5\) multiplies uncertainties by \(28\) on
\(\Imag\dt\) and \(25\) on \(\Real\dg\), leaving the uncertainty on \(\Real\dt\)
unchanged.
Figure~\ref{fig:scan} shows the marginal uncertainties for
\(\{\Real\dt,\Imag\dt,\Real\dg,\Imag\dg\}\) versus the \emph{total} off-peak luminosity, divided equally among four points at
\(\sqrt s=m_Z\pm1.7,\pm3.5~\GeV\) (\(1.25\%\) each at the \(5\%\)
benchmark).
The near degeneracy weakens with about \(1\%\) off peak;
the broad optimum spans \(4\)--\(9\%\). At \(5\%\), the two-dipole fit
leaves the uncertainty on \(\Real d_\tau^W\) unchanged
and increases it by a factor \(2.2\) on \(\Imag d_\tau^W\):
\begin{equation}
  \delta\Real\dt=4.5\times10^{-7},\quad
  \delta\Imag\dt=2.7\times10^{-6}
  \label{eq:final}
\end{equation}
at FCC-ee, and \(5.5\times10^{-7}\), \(3.2\times10^{-6}\) at CEPC.
The \((40,70,40)\,\mathrm{ab}^{-1}\) split of
Ref.~\cite{Janot:2015gjr}, \(53\%\) off peak, costs \(30\%\) here and gains
\(3.5\) on \(\Imag\dg\).
The fit also measures \(d_\tau^\gamma(m_Z^2)\) (End Matter).
These projections assume constant form factors across the scan and ideal
angular reconstruction.

\textit{Systematics.}---\(\langle\Ocal\rangle\)
is normalized by the observed count, so luminosity, \(B_Z\),
\(B_\pi\) and the overall efficiency cancel identically.  With \(W_0\) the SM distribution,
\(W_0\epsilon_{\rm even}\Ocal\) is CP odd and integrates to zero, so \emph{any} acceptance
satisfying \(\epsilon_+(\hat\ell)=\epsilon_-(-\hat\ell)\) produces no spurious mean.  CP-even acceptance errors miscalibrate the response but generate no signal.

CP-odd acceptance asymmetries bias transverse moments.  With \(\delta\langle\Ocal_4\rangle=6.2\times10^{-6}\) at FCC-ee,
a spurious up--down charge asymmetry \(\delta\) enters
\(\langle\Ocal_4\rangle\) as \(\delta\langle\mathrm{sgn}\cos\theta\rangle/3\)
with \(\langle\mathrm{sgn}\cos\theta\rangle=A_{\rm FB}^\tau\simeq0.017\), so a
\(\theta\)-independent asymmetry is suppressed by \(59\) and requires
\(\delta\lesssim10^{-3}\); an asymmetry correlated with the forward--backward
hemisphere is unsuppressed and requires \(\delta\lesssim2\times10^{-5}\).  The production-angle weight gives the factor \(13.5\) in Eq.~\eqref{eq:gain}
and suppresses the leading detector systematic by \(59\).

The CP-even rate constrains
\(|\dt|^2\) through \(A=908.2\) without determining a sign or phase.  Its
statistics-only radius, \(\rho_C=[z_C/(A\sqrt{N_0})]^{1/2}
=1.7\times10^{-4}\), requires precise rate normalization.  With a
fractional normalization uncertainty \(\delta_N\) the radius is
\(\rho_{\rm sys}=\sqrt{\delta_N/A}\), independent of sample size.  Branching fractions alone imply
\(\delta_N\gtrsim1\%\) and hence \(3.3\times10^{-3}\); normalizing inclusive
\(Z\to\tau\tau\) to \(Z\to\mu\mu\) gives the universality precision, \(\delta_N\simeq0.3\%\) and
\(1.8\times10^{-3}\), still weaker than ALEPH.  Reaching \(\rho_C\)
requires \(\delta_N=2.6\times10^{-5}\); matching Eq.~\eqref{eq:final}
requires \(\delta_N=1.8\times10^{-10}\).  Since
\(\rho_{\rm sys}\propto\sqrt{\delta_N}\), even a hundredfold normalization improvement
reduces the radius only tenfold; the rate contours in
Fig.~\ref{fig:reach}(a) show the gap.

\textit{Conclusion.}---Spin information determines how effectively a
\(Z\) sample tests CP violation. The longitudinal and triple-product
moments carry the helicity suppression \(2m_\tau/m_Z\)
and the vector-charge suppression \(|v_\tau/a_\tau|\).
Selected transverse moments avoid both; the sign-weighted normal moment
\(\Ocal_4\) retains the undiluted axial charge and enhances the response by \(521\)
relative to the triple product. The joint fit, Eq.~\eqref{eq:final}, improves
on ALEPH's published statistical errors by three orders of magnitude
using transverse moments and determines the timelike
electromagnetic dipole.
Building on Bernreuther and
Nachtmann~\cite{Bernreuther:1989kc,Bernreuther:1993nd} and the asymmetries
of Ref.~\cite{Vidal:1998jc}, we quantify the response hierarchy and
the axial contribution recovered by the production-angle weight,
identifying sensitive single angular moments for null tests.

Three experimental effects require further study:
tau-direction reconstruction dilutes the signal; initial-state radiation
smears \(q^2\) and rotates the \(\gamma/Z\) phase; and a CP-violating
phase in \(\tau\to\pi\nu\) is indistinguishable within the production
formalism. The radiation effect is calculable and small: an energy
loss \(\delta_E\) rotates the ratio in Eq.~\eqref{eq:propratio} by
\(\arctan(2\delta_E/\Gamma_Z)\), or \(4.6^\circ\) at
\(\delta_E=100~\mathrm{MeV}\), compared with \(70^\circ\) between the peak
and \(\sqrt s=m_Z\pm3.5~\GeV\). Radiation weakens but preserves the separation. A decay phase would make the extracted
dipoles depend on the decay mode, testable with \(\rho\nu\) and \(a_1\nu\); reconstruction
requires detector simulation.

\begin{acknowledgments}
The authors thank Manqi Ruan and Jia Liu for useful discussions.
This work is supported in part by the National Key Research and Development
Program of China under Grant No.~2020YFC2201501, by the Fundamental Research
Funds for the Central Universities, and by the National Natural Science
Foundation of P.R.~China (No.~12090064, 12375088, 12575096 and W2441004).
\end{acknowledgments}

\bibliographystyle{utphys}
\bibliography{ref}

@article{Yu:2019oal,
    author = "Yu, Dan and Ruan, Manqi and Boudry, Vincent and Videau, Henri and Brient, Jean-Claude",
    title = "{Higgs to $\tau\tau$ analysis in the future $e^{+}e^{-}$ Higgs factories}",
    eprint = "1903.12327",
    archivePrefix = "arXiv",
    primaryClass = "hep-ex",
    month = "3",
    year = "2019"
}

@article{Tsai:1971vv,
    author = "Tsai, Yung-Su",
    title = "{Decay Correlations of Heavy Leptons in e+ e- ---{\ensuremath{>}} Lepton+ Lepton-}",
    reportNumber = "SLAC-PUB-0932",
    doi = "10.1103/PhysRevD.13.771",
    journal = "Phys. Rev. D",
    volume = "4",
    pages = "2821",
    year = "1971",
    note = "[Erratum: Phys.Rev.D 13, 771 (1976)]"
}

@article{Bernreuther:1993nd,
    author = "Bernreuther, W. and Nachtmann, O. and Overmann, P.",
    title = "{The CP violating electric and weak dipole moments of the tau lepton from threshold to 500-GeV}",
    reportNumber = "HD-THEP-92-60",
    doi = "10.1103/PhysRevD.48.78",
    journal = "Phys. Rev. D",
    volume = "48",
    pages = "78--88",
    year = "1993"
}

@article{Bernabeu:1994wh,
    author = "Bernabeu, J. and Gonzalez-Sprinberg, G. A. and Tung, M. and Vidal, J.",
    title = "{The Tau weak magnetic dipole moment}",
    eprint = "hep-ph/9411289",
    archivePrefix = "arXiv",
    reportNumber = "FTUV-94-43, IFIC-94-38",
    doi = "10.1016/0550-3213(94)00525-J",
    journal = "Nucl. Phys. B",
    volume = "436",
    pages = "474--486",
    year = "1995"
}

@article{Belle:2002nla,
    author = "Inami, K. and others",
    collaboration = "Belle",
    title = "{Search for the electric dipole moment of the tau lepton}",
    eprint = "hep-ex/0210066",
    archivePrefix = "arXiv",
    reportNumber = "BELLE-2002-35",
    doi = "10.1016/S0370-2693(02)02984-2",
    journal = "Phys. Lett. B",
    volume = "551",
    pages = "16--26",
    year = "2003"
}

@article{Belle:2021ybo,
    author = "Inami, K. and others",
    collaboration = "Belle",
    title = "{An improved search for the electric dipole moment of the $\tau$ lepton}",
    eprint = "2108.11543",
    archivePrefix = "arXiv",
    primaryClass = "hep-ex",
    reportNumber = "Belle Preprint 2021-22, KEK Preprint 2021-26",
    doi = "10.1007/JHEP04(2022)110",
    journal = "JHEP",
    volume = "04",
    pages = "110",
    year = "2022"
}

@article{Buttazzo:2026amk,
    author = "Buttazzo, Dario and Levati, Gabriele and Ma, Yang and Maltoni, Fabio and Paradisi, Paride and Wang, ZeQiang",
    title = "{Probing $\tau$ lepton dipole moments at future Lepton Colliders}",
    eprint = "2604.14281",
    archivePrefix = "arXiv",
    primaryClass = "hep-ph",
    reportNumber = "IRMP-CP3-26-07, CERN-TH-2026-089, COMETA-2026-06",
    month = "4",
    year = "2026"
}

@article{Lu:2025hwy,
    author = "Lu, Peng-Cheng and Si, Zong-Guo and Zhang, Han and Zhang, Xin-Yi",
    title = "{Study of the {\ensuremath{\gamma}}{\ensuremath{\gamma}}{\textrightarrow}{\ensuremath{\tau}}+{\ensuremath{\tau}}- process including {\ensuremath{\tau}}+{\ensuremath{\tau}}- spin information in Pb-Pb ultraperipheral collisions and at a lepton collider}",
    eprint = "2511.18935",
    archivePrefix = "arXiv",
    primaryClass = "hep-ph",
    doi = "10.1103/2g5j-k778",
    journal = "Phys. Rev. D",
    volume = "113",
    number = "3",
    pages = "L031902",
    year = "2026"
}

@article{ALEPH:2002kbp,
    author = "Heister, A. and others",
    collaboration = "ALEPH",
    title = "{Search for anomalous weak dipole moments of the tau lepton}",
    eprint = "hep-ex/0209066",
    archivePrefix = "arXiv",
    reportNumber = "CERN-EP-2002-068",
    doi = "10.1140/epjc/s2003-01286-1",
    journal = "Eur. Phys. J. C",
    volume = "30",
    pages = "291--304",
    year = "2003"
}

@article{He:2025ewk,
    author = "He, Xiao-Gang and Liu, Chia-Wei and Ma, Jian-Ping and Yang, Chang and Zou, Zi-Yue",
    title = "{Precise measurement of CP violating $\tau$ EDM through $e^+e^- \to \gamma^*,\psi(2s) \to \tau^+\tau^-$}",
    eprint = "2501.06687",
    archivePrefix = "arXiv",
    primaryClass = "hep-ph",
    doi = "10.1007/JHEP04(2025)001",
    journal = "JHEP",
    volume = "04",
    pages = "001",
    year = "2025"
}

@article{Huang:2025ghw,
    author = "Huang, Zhong-Lv and Du, Xin-Yu and He, Xiao-Gang and Liu, Chia-Wei and Zou, Zi-Yue",
    title = "{Complex $\tau$ Electric Dipole Moment from GeV-Scale New Physics}",
    eprint = "2510.23348",
    archivePrefix = "arXiv",
    primaryClass = "hep-ph",
    doi = "10.1088/0256-307X/43/3/030201",
    journal = "Chin. Phys. Lett.",
    volume = "43",
    number = "3",
    pages = "030201",
    year = "2026"
}

@article{Du:2026yoi,
    author = "Du, Xin-Yu and He, Xiao-Gang and Huang, Zhong-Lv and Liu, Chia-Wei and Zou, Zi-Yue",
    title = "{Probing the imaginary parts and their $q^2$ dependences for the tau $g-2$ and EDM}",
    eprint = "2606.01178",
    archivePrefix = "arXiv",
    primaryClass = "hep-ph",
    month = "5",
    year = "2026"
}

@article{Ashby-Pickering:2022umy,
    author = "Ashby-Pickering, Rachel and Barr, Alan J. and Wierzchucka, Agnieszka",
    title = "{Quantum state tomography, entanglement detection and Bell violation prospects in weak decays of massive particles}",
    eprint = "2209.13990",
    archivePrefix = "arXiv",
    primaryClass = "quant-ph",
    doi = "10.1007/JHEP05(2023)020",
    journal = "JHEP",
    volume = "05",
    pages = "020",
    year = "2023",
    note = "[Erratum: JHEP 03, 166 (2026)]"
}

@article{Han:2025ewp,
    author = "Han, Tao and Low, Matthew and Su, Youle",
    title = "{Entanglement and Bell nonlocality in $\tau^+\tau^-$ at the BEPC}",
    eprint = "2501.04801",
    archivePrefix = "arXiv",
    primaryClass = "hep-ph",
    reportNumber = "PITT-PACC-2412",
    doi = "10.1007/JHEP10(2025)217",
    journal = "JHEP",
    volume = "10",
    pages = "217",
    year = "2025"
}

@article{Ai:2025wnt,
    author = "Ai, Tengyu and Bi, Qi and He, Yuxin and Liu, Jia and Wang, Xiao-Ping",
    title = "{Ultimate Quantum Precision Limit at Colliders: Conditions and Case Studies}",
    eprint = "2506.10673",
    archivePrefix = "arXiv",
    primaryClass = "hep-ph",
    reportNumber = "CPTNP-2025-019",
    doi = "10.1103/3m4t-pk9b",
    journal = "Phys. Rev. Lett.",
    volume = "135",
    number = "24",
    pages = "241804",
    year = "2025"
}

@article{Ai:2026udw,
    author = "Ai, Tengyu and Bi, Qi and He, Yuxin and Li, Zekun and Liu, Jia and Wang, Xiao-Ping",
    title = "{Collider Spin Tomography with Missing Neutrinos}",
    eprint = "2607.28346",
    archivePrefix = "arXiv",
    primaryClass = "hep-ph",
    reportNumber = "CPTNP-2026-020",
    month = "7",
    year = "2026"
}

@article{ParticleDataGroup:2024cfk,
    author = "Navas, S. and others",
    collaboration = "Particle Data Group",
    title = "{Review of particle physics}",
    doi = "10.1103/PhysRevD.110.030001",
    journal = "Phys. Rev. D",
    volume = "110",
    number = "3",
    pages = "030001",
    year = "2024"
}

@article{FCC:2025lpp,
    author = "Benedikt, M. and others",
    collaboration = "FCC",
    title = "{Future Circular Collider Feasibility Study Report: Volume 1, Physics, Experiments, Detectors}",
    eprint = "2505.00272",
    archivePrefix = "arXiv",
    primaryClass = "hep-ex",
    reportNumber = "CERN-FCC-PHYS-2025-0002",
    doi = "10.1140/epjc/s10052-025-15077-x",
    journal = "Eur. Phys. J. C",
    volume = "85",
    number = "12",
    pages = "1468",
    year = "2025",
    note = "[Erratum: Eur.Phys.J.C 86, 844 (2026)]"
}

@article{CEPCStudyGroup:2023quu,
    author = "Abdallah, Waleed and others",
    collaboration = "CEPC Study Group",
    title = "{CEPC Technical Design Report: Accelerator}",
    eprint = "2312.14363",
    archivePrefix = "arXiv",
    primaryClass = "physics.acc-ph",
    reportNumber = "IHEP-CEPC-DR-2023-01, IHEP-AC-2023-01",
    doi = "10.1007/s41605-024-00463-y",
    journal = "Radiat. Detect. Technol. Methods",
    volume = "8",
    number = "1",
    pages = "1--1105",
    year = "2024",
    note = "[Erratum: Radiat.Detect.Technol.Methods 9, 184--192 (2025)]"
}

@article{Ai:2024nmn,
    author = "Ai, Xiaocong and others",
    title = "{Flavor Physics at the CEPC: a General Perspective}",
    eprint = "2412.19743",
    archivePrefix = "arXiv",
    primaryClass = "hep-ex",
    doi = "10.1088/1674-1137/adf1f0",
    journal = "Chin. Phys. C",
    volume = "49",
    number = "10",
    pages = "103003",
    year = "2025"
}

@article{Ma:2023yvd,
    author = "Ma, Kai and Li, Tong",
    title = "{Testing Bell inequality through $h\to\tau\tau$ at CEPC*}",
    eprint = "2309.08103",
    archivePrefix = "arXiv",
    primaryClass = "hep-ph",
    doi = "10.1088/1674-1137/ad62d8",
    journal = "Chin. Phys. C",
    volume = "48",
    number = "10",
    pages = "103105",
    year = "2024"
}

@article{ALEPH:2005ab,
    author = "Schael, S. and others",
    collaboration = "ALEPH, DELPHI, L3, OPAL, SLD, LEP Electroweak Working Group, SLD Electroweak Group, SLD Heavy Flavour Group",
    title = "{Precision electroweak measurements on the $Z$ resonance}",
    eprint = "hep-ex/0509008",
    archivePrefix = "arXiv",
    primaryClass = "hep-ex",
    reportNumber = "CERN-PH-EP-2005-041, SLAC-R-774, LEPEWWG-2005-01",
    doi = "10.1016/j.physrep.2005.12.006",
    journal = "Phys. Rept.",
    volume = "427",
    pages = "257--454",
    year = "2006"
}

@article{Diehl:1993br,
    author = "Diehl, M. and Nachtmann, O.",
    title = "{Optimal observables for the measurement of three gauge boson couplings in $e^+ e^- \to W^+ W^-$}",
    reportNumber = "HD-THEP-93-37",
    doi = "10.1007/BF01555899",
    journal = "Z. Phys. C",
    volume = "62",
    pages = "397--412",
    year = "1994"
}

@article{Bernreuther:1989kc,
    author = "Bernreuther, W. and Nachtmann, O.",
    title = "{CP Violating Correlations in Electron Positron Annihilation Into $\tau$ Leptons}",
    reportNumber = "LBL-27398",
    doi = "10.1103/PhysRevLett.63.2787",
    journal = "Phys. Rev. Lett.",
    volume = "63",
    pages = "2787",
    year = "1989",
    note = "[Erratum: Phys.Rev.Lett. 64, 1072 (1990)]"
}

@article{Vidal:1998jc,
    author = "Vidal, J. and Bernabeu, J. and Gonzalez-Sprinberg, G.",
    title = "{Tau weak dipole moments from azimuthal asymmetries}",
    eprint = "hep-ph/9812373",
    archivePrefix = "arXiv",
    primaryClass = "hep-ph",
    doi = "10.1016/S0920-5632(99)00471-5",
    journal = "Nucl. Phys. B Proc. Suppl.",
    volume = "76",
    pages = "221--228",
    year = "1999"
}

@article{Bernabeu:2004ww,
    author = "Bernabeu, J. and Gonzalez-Sprinberg, G. A. and Vidal, J.",
    title = "{CP violation and electric-dipole-moment at low energy tau-pair production}",
    eprint = "hep-ph/0404185",
    archivePrefix = "arXiv",
    primaryClass = "hep-ph",
    doi = "10.1016/j.nuclphysb.2004.08.038",
    journal = "Nucl. Phys. B",
    volume = "701",
    pages = "87--102",
    year = "2004"
}

@article{Bernabeu:2006wf,
    author = "Bernabeu, J. and Gonzalez-Sprinberg, G. A. and Vidal, J.",
    title = "{CP violation and electric-dipole-moment at low energy tau production with polarized electrons}",
    eprint = "hep-ph/0610135",
    archivePrefix = "arXiv",
    primaryClass = "hep-ph",
    doi = "10.1016/j.nuclphysb.2006.11.023",
    journal = "Nucl. Phys. B",
    volume = "763",
    pages = "283--292",
    year = "2007"
}

@article{Sun:2024vcd,
    author = "Sun, Xulei and Wu, Yongcheng and Zhou, Xiaorong",
    title = "{Sensitivity Study of the Tau Lepton Electric Dipole Moment at the Super Tau-Charm Facility}",
    eprint = "2411.19469",
    archivePrefix = "arXiv",
    primaryClass = "hep-ph",
    doi = "10.1088/1674-1137/adf6e0",
    journal = "Chin. Phys. C",
    volume = "49",
    number = "11",
    pages = "113001",
    year = "2025"
}

@article{Sanchez:1997kp,
    author = "Sanchez, F.",
    title = "{Optimal observable to measure the imaginary component of the tau anomalous weak-magnetic dipole moment}",
    doi = "10.1016/S0370-2693(97)01051-4",
    journal = "Phys. Lett. B",
    volume = "412",
    pages = "137--142",
    year = "1997"
}

@article{Chupp:2017rkp,
    author = "Chupp, T. E. and Fierlinger, P. and Ramsey-Musolf, M. J. and Singh, J. T.",
    title = "{Electric dipole moments of atoms, molecules, nuclei, and particles}",
    eprint = "1710.02504",
    archivePrefix = "arXiv",
    primaryClass = "physics.atom-ph",
    doi = "10.1103/RevModPhys.91.015001",
    journal = "Rev. Mod. Phys.",
    volume = "91",
    number = "1",
    pages = "015001",
    year = "2019"
}

@article{Roussy:2022cmp,
    author = "Roussy, Tanya S. and Caldwell, Luke and Wright, Trevor and Cairncross, William B. and Shagam, Yuval and Ng, Kia Boon and Schlossberger, Noah and Park, Sun Yool and Wang, Anzhou and Ye, Jun and Cornell, Eric A.",
    title = "{An improved bound on the electron's electric dipole moment}",
    eprint = "2212.11841",
    archivePrefix = "arXiv",
    primaryClass = "physics.atom-ph",
    doi = "10.1126/science.adg4084",
    journal = "Science",
    volume = "381",
    number = "6653",
    pages = "46--50",
    year = "2023"
}

@article{Altakach:2022ywa,
    author = "Altakach, Mohammad Mahdi and Lamba, Priyanka and Maltoni, Fabio and Mawatari, Kentarou and Sakurai, Kazuki",
    title = "{Quantum information and CP measurement in $H \to \tau^+\tau^-$ at future lepton colliders}",
    eprint = "2211.10513",
    archivePrefix = "arXiv",
    primaryClass = "hep-ph",
    doi = "10.1103/PhysRevD.107.093002",
    journal = "Phys. Rev. D",
    volume = "107",
    number = "9",
    pages = "093002",
    year = "2023"
}

@article{Davier:1992nw,
    author = "Davier, M. and Duflot, L. and Le Diberder, F. and Rouge, A.",
    title = "{The Optimal method for the measurement of tau polarization}",
    doi = "10.1016/0370-2693(93)90101-M",
    journal = "Phys. Lett. B",
    volume = "306",
    pages = "411--417",
    year = "1993"
}

@article{Janot:2015gjr,
    author = "Janot, Patrick",
    title = "{Direct measurement of $\alpha_{QED}(m^2_{\rm Z})$ at the FCC-ee}",
    eprint = "1512.05544",
    archivePrefix = "arXiv",
    primaryClass = "hep-ph",
    doi = "10.1007/JHEP02(2016)053",
    journal = "JHEP",
    volume = "02",
    pages = "053",
    year = "2016",
    note = "[Erratum: JHEP 11, 164 (2017)]"
}

@article{Kosnik:2026wdm,
    author = "Ko\v{s}nik, Nejc and Polonsky, Zachary and Smolkovi\v{c}, Aleks",
    title = "{Current and future constraints on heavy New Physics from $\tau$ weak dipole moments}",
    eprint = "2606.07232",
    archivePrefix = "arXiv",
    primaryClass = "hep-ph",
    year = "2026"
}

@article{Pospelov:2013sca,
    author = "Pospelov, Maxim and Ritz, Adam",
    title = "{CKM benchmarks for electron electric dipole moment experiments}",
    eprint = "1311.5537",
    archivePrefix = "arXiv",
    primaryClass = "hep-ph",
    doi = "10.1103/PhysRevD.89.056006",
    journal = "Phys. Rev. D",
    volume = "89",
    number = "5",
    pages = "056006",
    year = "2014"
}

@article{Huet:1994jb,
    author = "Huet, Patrick and Sather, Eric",
    title = "{Electroweak Baryogenesis and Standard Model CP Violation}",
    eprint = "hep-ph/9404302",
    archivePrefix = "arXiv",
    doi = "10.1103/PhysRevD.51.379",
    journal = "Phys. Rev. D",
    volume = "51",
    pages = "379--394",
    year = "1995"
}
\clearpage

\onecolumngrid
\vspace{1em}
\begin{center}\textbf{\large End Matter}\end{center}
\vspace{0.5em}
\twocolumngrid

\textit{Angular distribution and Fisher information.}---The normalized
visible distribution is
\(W=W_0+\Real\dt\,\widehat W_R+\Imag\dt\,\widehat W_I+\Ocal(|\dt|^2)\)
with \(\dd x=\dd c\,\dd c_-\dd\phi_-\dd c_+\dd\phi_+\),
\(c=\cos\theta\), \(c_\mp=\cos\theta_\mp\), \(s_\mp=\sin\theta_\mp\), all
angles in the triad of Eq.~\eqref{eq:triad}, and
\(\int\!\dd x\,W_0=1\), \(\int\!\dd x\,\widehat W_{R,I}=0\).  Defining
\(\mathcal{K}_1=a_\tau\beta c+v_\tau A_e\) and
\(\mathcal{K}_2=v_\tau c+a_\tau\beta A_e\), where \(A_e\) is the electron
asymmetry parameter of Eq.~\eqref{eq:O3resp}, we obtain
\begin{align}
  \widehat W_R&=\frac{3\beta s}{32\pi^2\Delta_0}\frac{m_Z}{2m_\tau}\Big\{
   -\mathcal{K}_1\left(s_-\sin\phi_-+s_+\sin\phi_+\right) \notag\\
  &\quad-v_\tau\sqrt{1-\beta^2}\,s\,s_-s_+\sin(\phi_+-\phi_-) \notag\\
  &\quad-\mathcal{K}_2\left(s_-\sin\phi_-c_+-s_+\sin\phi_+c_-\right)\Big\},
  \label{eq:WR}\\
  \widehat W_I&=\frac{3\beta s}{32\pi^2\Delta_0}\frac{m_Z}{2m_\tau}\Big\{
   -\mathcal{K}_2\left(s_-\cos\phi_-+s_+\cos\phi_+\right) \notag\\
  &\quad-v_\tau\sqrt{1-\beta^2}\,s\left(c_-+c_+\right) \notag\\
  &\quad+\mathcal{K}_1\left(s_+\cos\phi_+c_--s_-\cos\phi_-c_+\right)\Big\}.
  \label{eq:WI}
\end{align}
The response hierarchy follows directly from
Eqs.~\eqref{eq:WR} and \eqref{eq:WI}.  The overall \(m_Z/2m_\tau\) gives the
dipole enhancement, while the second lines, which feed
\(\Ocal_1\) and \(\Ocal_2\), contain the compensating factor
\(\sqrt{1-\beta^2}\).  The first line of \(\widehat W_R\) carries the
\(a_\tau\beta c\) term of \(\mathcal{K}_1\), odd in \(c\); the moment \(\Ocal_4\)
retains this term, which cancels under unweighted symmetric \(c\) integration, while the
\(v_\tau A_e\) term survives without the weight.

The scores \(\widehat W_{R,I}/W_0\) are the optimal observables of
Refs.~\cite{Bernreuther:1993nd,Davier:1992nw,Diehl:1993br}.
The per-event Fisher matrix
\(\Ical_{ij}=\int\!\dd x\,\widehat W_i\widehat W_j/W_0\) has
\(\Ical_{RI}=0\) identically, since
\((\phi_-,\phi_+)\mapsto(-\phi_-,-\phi_+)\) makes \(\widehat W_R\) odd and
\(W_0,\widehat W_I\) even.  Numerically, at the couplings of
Eq.~\eqref{eq:couplings} with full acceptance,
\begin{equation}
  \Ical_{RR}=549.7,\ \ \Ical_{II}=206.1
  \quad(\pi^-\pi^+),
  \label{eq:fisherpipi}
\end{equation}
and, after integrating over the opposite decay,
\begin{equation}
  \Ical^{(1)}_{RR}=174.6,\ \ \Ical^{(1)}_{II}=23.2
  \quad(\text{one analyzed pion}),
  \label{eq:fisher1}
\end{equation}
These give \(\delta\Real\dt\geq7.6\times10^{-2}/\sqrt N\) and
\(\delta\Imag\dt\geq2.1\times10^{-1}/\sqrt N\) for \(N\) independent
single-pion measurements. The FCC-ee and CEPC Cram\'er--Rao rows of
Table~\ref{tab:reach} use \(N=2N_{\rm eff}^\mp\); the ALEPH row uses
Eq.~\eqref{eq:fisherpipi} with \(N=1901\) double-pion events.

These are classical bounds for the specified angular distributions at the SM point.
A joint treatment of events with both pion decays reconstructed can access
additional information through their correlations.
These bounds do not establish the quantum precision limit of the produced
tau pair, which need not be attainable from decay momenta~\cite{Ai:2025wnt}.
As a cross-check, ALEPH's published statistical errors for the \(1901\)
candidates exceed the double-pion bound by \(1.60\) and \(1.63\), consistent with
finite acceptance in an optimal-observable analysis.

Since \(\delta\theta_i\propto\Ical_{ii}^{-1/2}\), the \(8\%\) and \(14\%\)
uncertainty gaps mean that \(\Ocal_4\) and \(\Ocal_3\) retain \(86\%\) and \(78\%\)
of \(\Ical^{(1)}_{RR}\) and \(\Ical^{(1)}_{II}\), respectively.
Calculated from the full single-pion angular distribution, the Cram\'er--Rao bound
sets a lower uncertainty limit for unbiased estimators when pion measurements
are treated independently, regardless of the chosen moment.
It therefore gauges how efficiently each moment extracts the available single-pion information.
The small gaps indicate nearly optimal precision within this analysis.

The optimal-observable construction is not
new~\cite{Bernreuther:1993nd,Davier:1992nw,Diehl:1993br}, and was applied to
the tau weak-\emph{magnetic} dipole at LEP~\cite{Bernabeu:1994wh,Sanchez:1997kp}.
The new result is the classical Fisher information for the weak \emph{electric} dipole in this channel, Eqs.~\eqref{eq:fisherpipi} and \eqref{eq:fisher1}: it
provides a common benchmark that attributes the losses in \(\Ocal_1\) and \(\Ocal_2\)
to the chosen moments.

\textit{Cross-checks and the low-energy limit.}---The moments in Eqs.~\eqref{eq:O3}--\eqref{eq:O4} probe the transverse polarizations studied in Ref.~\cite{Vidal:1998jc}, whose
normal-spin coefficient is proportional to
\(\gamma\beta\sin\theta\,[\,2v_\tau v_ea_e
+(v_e^2+a_e^2)a_\tau\beta\cos\theta\,]\), with
\(\gamma=m_Z/2m_\tau\).  The bracket shows the role of the weight in
Eq.~\eqref{eq:O4}: the first term is even in \(\cos\theta\) and
the second, larger by \(a_\tau\beta/(v_\tau A_e)=92.2\), is odd and is retained only with
the sign weight.  After angular integration, their ratio,
\(\tfrac{4}{3\pi}a_\tau\beta/(v_\tau A_e)=39.14\), agrees with our
numerical gain \(39.15\) within \(0.03\%\).  Consistently, the LEP-era sensitivity of order
\(2\times10^{-4}\) quoted in that work matches Eq.~\eqref{eq:O4resp} scaled to the
\(1.7\times10^{7}\) \(Z\) decays of LEP, \(2.5\times10^{-4}\), and not
the unweighted asymmetry, \(1.0\times10^{-2}\).

Changing the input scheme gives a \(2\%\) effect: taking \(G_F\) rather than
\(\alpha\) as the third input raises both charges by
\([1/(1-\Delta r)]^{1/2}=1.0185\).  Because the dipole term of
Eq.~\eqref{eq:vertexZ} carries no tau charge, a larger charge dilutes the
dipole-to-SM ratio: every \(\dt\) reach quoted above weakens by \(1.0185\),
and every \(\dg\) reach by its square.
It leaves \(A_\tau\) and \(a_\tau/v_\tau\) untouched, since
\((G_F,m_W,m_Z)\) and \((\alpha,m_W,m_Z)\) both take \(s_W^2\) from
\(m_W/m_Z\) and differ only in the overall normalization; the shift that
brings \(A_\tau\) onto \(A_\tau^{\rm exp}\) is a loop effect, carried here by
\(s^2_{\rm eff}\).

This hierarchy requires \(\beta\to1\). At tau-charm energies, photon exchange
dominates and \(\beta\) is small, reversing the ordering: at \(\sqrt s=4.26~\GeV\),
the longitudinal electromagnetic-dipole response is \(0.273\), compared
with \(0.164\) for the production-angle-weighted transverse asymmetry.
The normal asymmetry vanishes by parity conservation in pure photon
exchange; at \(\sqrt s=3.77~\GeV\), the longitudinal and weighted transverse responses are \(0.154\) and \(0.082\).
Thus \(\Ocal_1\) and \(\Ocal_2\) are appropriate for BESIII and the Super Tau-Charm
Facility~\cite{Bernabeu:2004ww,Bernabeu:2006wf,Sun:2024vcd,Belle:2002nla,%
Belle:2021ybo,He:2025ewk}; their suppression appears at the \(Z\) pole,
where the response ratio differs by two orders of magnitude.

ALEPH's full differential fit retained the transverse information and
avoided the penalties of Eqs.~\eqref{eq:O3-over-O1} and \eqref{eq:gain}. With \(4\times10^6\) \(Z\) decays,
the analysis was statistics limited and the moment choice immaterial;
at \(6\times10^{12}\) \(Z\) decays, that choice matters. Simple angular moments
permit analytic acceptance corrections and the symmetry-protected null
described below. An optimal observable requires event-by-event evaluation against simulated \(W_0\),
inheriting the simulation's systematic uncertainties.

\textit{Electromagnetic dipole sensitivity.}---The joint fit determines \(\dg\) simultaneously with the weak dipole.  At the \(5\%\) baseline, the marginal errors are
\begin{equation}
  \delta\Real\dg=9.6\times10^{-6},\quad
  \delta\Imag\dg=2.1\times10^{-5}
  \label{eq:gammareach}
\end{equation}
at FCC-ee, and \(1.2\times10^{-5}\), \(2.6\times10^{-5}\) at CEPC. At FCC-ee, the peak-only errors are \(1.2\times10^{-4}\) and
\(8.5\times10^{-4}\): the scan improves both photon-dipole uncertainties by more than an order of
magnitude, as shown in Fig.~\ref{fig:reach}(b). The dispersive (absorptive)
uncertainty remains \(20\) (\(8\)) times larger than its weak counterpart,
consistent with the propagator suppression in Eq.~\eqref{eq:propratio}.

For \(\Real d_\tau^\gamma\), the projected precision is \(118\) times that of
the Belle bound in Ref.~\cite{Belle:2021ybo}, as listed in Table~\ref{tab:reach}.
These projections cannot be directly compared with the Belle
limits~\cite{Belle:2002nla,Belle:2021ybo}, which constrain the same form
factor at \(q^2=(10.58\,\GeV)^2\); the published limits are quoted in \(\ecm\)
and converted here using \(\ee\hbar c/(2m_\tau)=5.552\times10^{-15}\,\ecm\).
Equation~\eqref{eq:gammareach} refers to \(q^2=m_Z^2\), where the
absorptive part requires rescattering on top of CP violation and so vanishes
at tree level in the effective theory.  The two measurements are complementary, and the form factor's \(q^2\) dependence has recently been made explicit~\cite{Du:2026yoi}. Complementary spin observables have also been studied in
\(\gamma\gamma\to\tau^+\tau^-\)~\cite{Lu:2025hwy}.

\textit{Charge-resolved null tests.}---\(\Ocal_4\) has no Standard Model pedestal:
its mean vanishes in each charge sample, allowing separate checks.  By contrast \(\langle\hat\ell^{\,\mp}\!\cdot\hat z\rangle=\mp0.049\)
individually, so the null of \(\Ocal_1\) comes from canceling two charge means
\(8000\) times the target precision, requiring the two charge-dependent polarimetry scales to agree to \(1.3\times10^{-4}\).

\textit{Numerical method.}---We compute all responses from explicit helicity amplitudes with both propagators and both dipoles, using the
production density matrix and the decomposition
\(W\propto1+\bm P_-\!\cdot\hat\ell^{\,-}+\bm P_+\!\cdot\hat\ell^{\,+}
+\hat\ell^{\,-T}C\hat\ell^{\,+}\); setting \(D^\gamma/D^Z\to0\) reproduces
Eqs.~\eqref{eq:O1resp}--\eqref{eq:O5resp} to four digits, with all five CP-odd
SM means vanishing at the \(10^{-18}\) level.

\textit{Reconstruction and identifiability.}---With the visible measurement described by a coarse-grained positive
operator-valued measure (POVM) on the production density matrix~\cite{Ashby-Pickering:2022umy,Ma:2023yvd,%
Altakach:2022ywa,Han:2025ewp}, Ref.~\cite{Ai:2026udw} finds that in
\(\tau^+\tau^-\to\pi^+\pi^-\nu\bar\nu\) fourteen of the fifteen spin coefficients remain identifiable; the exception is \(C_{nr}-C_{rn}\), corresponding to
\(\Ocal_2\), since
\(\langle(\hat\ell^{\,-}\times\hat\ell^{\,+})\cdot\hat k_-\rangle\propto
C_{rn}-C_{nr}\).  The triple product is therefore helicity suppressed and inaccessible
without vertex information.  The transverse single-spin polarizations probed by \(\Ocal_3\) and \(\Ocal_4\) are among the
fourteen identified by template-free unfolding in Ref.~\cite{Ai:2026udw}.

\end{document}